# Oscillatory exchange bias and training effects in nanocrystalline $Pr_{0.5}Ca_{0.5}MnO_3$

S. Narayana Jammalamadaka[1*], S. S. Rao[2,3*], S.V.Bhat[3], J. Vanacken[1] and V. V. Moshchalkov[1]

[1]*INPAC – Institute for Nanoscale Physics and Chemistry, Nanoscale Superconductivity and Magnetism and Pulsed Fields Group, K.U. Leuven, Celestijnenlaan 200D, B–3001 Leuven, Belgium*

[2]*INPAC – Institute for Nanoscale Physics and Chemistry, Semiconductor Physics Laboratory, K.U. Leuven, Celestijnenlaan 200D, B–3001 Leuven, Belgium*

[3]Department of Physics, Indian Institute of Science, Bangalore – 560012, India

**Abstract**

We report on exchange bias effects in 10 nm particles of $Pr_{0.5}Ca_{0.5}MnO_3$ which appear as a result of competing interactions between the ferromagnetic (FM)/anti-ferromagnetic (AFM) phases. The fascinating new observation is the demonstration of the temperature dependence of oscillatory exchange bias (OEB) and is tunable as a function of cooling field strength below the SG phase, may be attributable to the presence of charge/spin density wave (CDW/SDW) in the AFM core of PCMO10. The pronounced training effect is noticed at 5 K from the variation of the EB field as a function of number of field cycles (n) upon the field cooling (FC) process**.** For n > 1, power-law behavior describes the experimental data well; however, the breakdown of spin configuration model is noticed at n ≥ 1.

*These authors contributed equally to this work

**Introduction**

In a ferromagnet (FM)/anti-ferromagnet (AFM) coupled system, the shift of hysteresis (MH) loops along the field axis after cooling the sample in the presence of magnetic field across the Neel temperature ($T_N$) of AFM material can be termed as "exchange bias (EB)"[1]. EB is believed to be accompanied by a gradual reduction in the EB field ($H_{EB}$) during the consecutive loops (n) and an enhancement of coercivity ($H_C$) within the same magnetic field range. The former is known as 'training effect' and is often observed in connection with a change in shape of MH loop, in addition to the reduction in $H_{EB}$ and enhancement in $H_C$[2]. This effect has been observed in many systems such as nanoparticles[3], FM/AFM multilayers[4] and disordered manganites[3], which is not mandatory for all EB systems[5] and has been extensively used in conjunction with EB effect in spintronic devices.[6,7]

The mixed valent manganites have attracted considerable interest due to their technological applications as magnetic sensors and recording devices[8]. These manganites having the general formula, $R_{1-x}A_xMnO_3$ ( where R is a rare-earth ion and A is a alkaline-earth ion) have been well investigated for their exotic electronic and magnetic properties[9,10], such as colossal magnetoresistance (CMR) and charge-ordered (CO) phase, to name a few. Recently it has been proved experimentally that at the nanoscale, these manganites can have a FM ground state and the CO phase[11] is absent. The magnetic properties of nanoscale manganites are fundamentally different from their bulk counterparts due to the significant enhancement of surface spins[3]. However, the assignment and attribution of the EB and subsequent consequences such as training effects have not been studied in detail particularly in the nanoscale $Pr_{0.5}Ca_{0.5}MnO_3$ (PCMO). Nevertheless, EB effects would help in understanding the exchange anisotropy coupling between FM and AFM clusters, and consequently can help to understand the CMR effect further. Hence, in the present paper we studied the EB and its training effect in 10 nm particles of $Pr_{0.5}Ca_{0.5}MnO_3$ (PCMO10). The present results would in turn helpful in broadening the current understanding of the EB systems, particularly on nanoscale CO manganites, and may have applications in spintronic devices.

The discussion related to the preparation, characterization and the first magnetization results obtained on PCMO10 have been reported[12,13] earlier. MH loops were measured using a vibrating sample magnetometer (Oxford instruments) in zero – field – cooling (ZFC) or field cooling (FC) conditions in the temperature (T) range from 5 – 40 K and at various magnetic field ranges.

In case of nano particles such as PCMO10, M(H) loop at moderate magnetic field does not saturate often. In particular, materials involving disordered magnetic and/or glassy magnetic phases or canted spin configuration or a system with large anisotropy do not show saturating trend even for H > 140 kOe. To test this on PCMO10 and in an effort to eliminate the contribution of minor loop effect while evaluating the EB, we had performed static high magnetic field (140 k Oe) isothermal M-H measurements at 5 K and the results have been disclosed in our earlier work[13]. As it can be immediately noticed (cf. see, Fig 2a in Ref.13), the "humming bird-like" unsaturated MH curve is observed. Thus, apparently higher magnetic field needs to be applied to detect saturated M(H) loop, crucial for investigating intrinsic EB effect. Albeit, many recent reports[14,15] have appeared on EB measured from unsaturated MH loops, because of very high magnetic fields are apparently necessary to saturate MH curve, inaccessible to many of the laboratories.

To gain further insights, we have extended the magnetic field range aiming to detect saturated M(H) curves. The high field magnetization (M-H) measurements were performed at various T's using the pulsed magnetic field facility at the University of Leuven. We have applied pulsed magnetic fields up to 170 kOe with pulse duration of about 20 ms by discharging a capacitor bank through a specially designed magnet coil. For M(H) measurements, we used the induction method by employing pickup coils, where the voltage induced in the pickup coils was integrated numerically to obtain the magnetization. At various T's, we collected M(H) data by sweeping the magnetic fields up to about 170 kOe [cf. see, Fig.3 in Ref.12]. At any T covered, the same "humming bird-like" unsaturated M(H) curve is observed, similar to the shape of the M(H) loop recorded up to 10 kOe (cf. see, Fig.1 in the current work). With more confidence, we have extended our magnetic field sweep range up to 300 kOe and have measured M(H) curve at 8 K [cf. see, Fig.2 in Ref.12]. With success, we find a saturated M(H) curve at around 250 kOe, indicative of global ferromagnetism (GFM). We believe that, in such circumstances, it is hard to exclude the minor loop effect in the evolution of EB within easily accessible magnetic field range. Much more sensitive experimental tools such as polarized neutron reflectometry (PNR) and x-ray magnetic circular dichroism (XMCD) may allow us to separate the minor loop effect contribution, which are inaccessible to us, as of now. Therefore, in the current work, the EB has been evaluated within the magnetic field range of -10 kOe to +10 kOe, though with unavoidable contribution of minor loop effect.

To investigate the EB effects in PCMO10, we measured MH loops from -10 to 10 kOe at various T's of 5, 8, 10, 13,15,17, 20,23, 25,28, 30,33, 35 and 38 K in ZFC and FC regimes. For each FC measurement the sample was warmed up to 300 K and cooled to the desired T in the presence of 10 kOe. The MH loops at few selected T's (5, 10, 15, 20, 25 and 30 K) are shown in Fig.1. As it can be readily noticed at 5 K, the MH-loop midpoints of the ZFC and FC curves differ; as the vertical midpoint shift is indicated by the dotted lines (for clarity dotted lines are not shown for the other temperatures). Horizontal and vertical loop shifts are noticed in the FC conditions. The horizontal shift can be termed as EB phenomena as a result of competing interactions between the FM/AFM aligned spins at the interfaces. The vertical loop shifts can be attributed to the uncompensated pinned spins at the FM/AFM interface.

All the data points gathered in Fig. 2a were measured by raising the sample temperature to 300 K after one field cooling. For example, the exchange bias at 5 K was measured after field cooling the sample down to 5 K under 10 kOe. Before we measure our next point, we raised the sample temperature to 300 K, then apply the field of 10 kOe while cooling down. This process has been repeated for all the temperatures.

Figure 2a shows the variation of $H_{EB}$ as a function of T for PCMO10. $H_{EB}$ is determined from the horizontal shift in the midpoint of the MH loop. The sign of the $H_{EB}$ remains unchanged throughout the whole T range. The most fascinating observation is that the T dependence of $H_{EB}$ is not monotonic but oscillatory, as observed below the SG freezing T of $T_{SG} = 40$ K [12]. Now we describe the various plausible physical processes that may account for OEB (T), beginning with the most significant one. As is well-known[16,17], in strongly correlated systems such as perovskite manganites, the charge, spin and orbital degrees of freedom are strongly coupled. Also, it has been established that[16,17], at half doping (x = 0.5), the AFM CO manganites are characterized by incommensurate charge density wave (ICDW), similar to the incommensurate spin density wave (ISDW) present in other FM/AFM layers[18,19], in which the wave length of incommensurate wave is strongly T dependent below $T_N = 175$ K of bulk PCMO. The presence of incommensurate charge density wave (ICDW) has been shown[20 - 24] to be present in a typical charge ordered polycrystalline manganites such as $LaCaMnO_3$, $PrCaMnO_3$ and $NdSrMnO_3$, in which, the charge ordered anti-ferromagnetic phase as well as ferromagnetic phases co-exist, and are expected to show 'exchange bias' similar to the sample being studied in the current work. In such materials, it has been shown that wavelength (wave vector) is a strong function of temperature in a strong magnetic back ground.

The obtained results may now be understood more clearly in terms of a phenomenological core-shell model[25], where the core of the PCMO10 shows AFM behavior and the surrounding shell behaves as a FM/SG-like system due to uncompensated surface spins. In the present work on PCMO10, the EB may arise from the uncompensated magnetic moments generated at the interface between the AFM core and FM shell arising due to finite size scaling effects. In the prior works on epitaxial Fe/Cr(001)[18] and (100)Cr/Py[19], the observation of OEB (T) has been reported to be due to the *"thermally driven wavelength expansion of an ISDW in Cr"*. Similarly, in the present case as well, we may speculate that the oscillations in EB may arise due to an incommensurate charge/spin density wave in the AFM core of PCMO10. Due to the expansion of wave length of incommensurate wave with increasing T, the net AFM moments at the core (AFM)-shell (FM) interface of PCMO10 would change its orientation from one direction to the opposite direction at T of $T_1$, $T_2$ and $T_3$. This would result in a change of sign in EB from $T_1$ to $T_2$, and from $T_2$ to $T_3$. Because of the roughness, grain boundaries and strain at the core (AFM)-shell (FM) interface in polycrystalline PCMO10, the sign change of EB is often obscured, whereas the oscillation in EB is more readily observable. For clear picture, the actual spin structure of AFM core region of PCMO10 needs to be studied with more sensitive experimental techniques.

Secondly, the observed OEB behaviour in PCMO10 might be related to its underlying magnetic phase diagram, similar to the reported[26] scenario in the case of Fe/Gd layers. Yet, in another work[27] on Fe/Cr bi-layers, the OEB (T) was explained by considering the interface alloying and the formation of a cluster spin-glass phase. The latter mechanism might also be relevant to the present case, as PCMO10 was shown[12] to exhibit SG phase at low T = 40 K.

A vertical shift of the MH curve, i. e. $|M(+H_0^{max})| \neq |M(-H_0^{max})|$ is clearly seen at all measured T's below 40 K (~ $T_{SG}$) (cf. see Fig1). Possibly, the net uncompensated spins induced at the FM/AFM interface could have contributed substantially to the observed $H_{EB}$. To verify this, the sample was cooled from 300 to 5 K at various field-cooling (FC) strengths in the range between 0.3 – 70 kOe. Fig. 2b shows the variation of magnetization shift ΔM **(abs(M(+H))-abs(M(-H)))** with FC. As shown in this Fig, the ΔM increases with increasing the cooling-field strength. The origin of this ΔM could be due to the uncompensated spins pinned at the FM/AFM interface. At this point it is worth mentioning that, similar to the present sample, an interfacial uncompensated AFM spins have been observed[3] in $Sm_{0.5}Ca_{0.5}MnO_3$ nanoparticles as well. While trying to unravel the uncompensated spins in our sample, similar to the experiment done by Huang and co-authors [28], a two-step field-cooling experiment has been performed. From our early ac – susceptibility measurements[12] on PCMO10, uncompensated pinned spins are present below 100 K, at around which the ferromagnetic phase begins to appear, and hence we first cooled the sample from 300 to 100 K in the presence of 10 kOe. At 100 K, we turned the magnetic field off and cooled the sample in zero field down to 5 K. We find that the observed MH loop was symmetric without any shift in magnetization (ΔM) and exchange bias observed as shown in Fig. 2c, thus supporting the presence of the uncompensated spins at the interface.

Concurrently, the observed huge magnetization shift ΔM ~ 1.6 $\mu_B$ (at 70 kOe FC) cannot be corroborated to the uncompensated spins alone at the interface of FM/AFM; instead there could be some other mechanisms by which additional spins become responsible for the change in ΔM.

Huge changes in the ΔM values were observed[28] in the ZnCoO/NiO system, and this change has been correlated with the existence of a SG-like phase in ZnCoO. As reported in Ref.12, in PCMO10 as well, there is a frequency dependent peak shift in real part of ac susceptibility data which confirms the SG behavior. The presence of such SG behaviour could cause to additional pinned mechanisms leading to vertical loop shift as well as huge change in ΔM.

Additional measurements (data not shown) were performed to check the OEB (T) under higher cooling field strength of 30 kOe and for second loop (n =2) with the cooling field strength of 10 kOe using SQUID magnetometry. In both cases, the exchange bias is found to go down linearly with the temperature, against our earlier observation - oscillatory behaviour for the cooling field strength (10 kOe) and for the first loop (n = 1). From the observed results we believe that there is a strong influence of the cooling – field strength and thermal cycling on the spins which exist at the FM/AFM interfaces resulting in the modification of the OEB(T) observed in the first loop and under the cooling field strength of 10 kOe.

Figure 2d illustrates the $H_{EB}$ behaviour as a function of cooling field ($H_{FC}$). It can be observed that, in low $H_{FC}$ the $H_{EB}$ increases and $H_{EB}$ diminishes for higher $H_{FC}$. As the $H_{FC}$ increases, the effective Zeeman energy increases and this can make more and more spins aligned until all spins are parallel to the external field, resulting in the 'depinning' of the pinned region. At higher $H_{FC}$, the $H_{EB}$ decreases due to the reduction of the pinned spins. The variation of $H_{EB}$ with $H_{FC}$ can be explained by the following relation.[24]

$$-H_{EB} \propto J_i \left[ \frac{J_i \mu_0}{(g\mu_B)^2} L\left(\frac{\mu H_{FC}}{k_B T_f}\right) + H_{FC} \right] \quad \ldots\ldots\ldots(1)$$

$J_i$ is the interface exchange coupling constant, g is the gyromagnetic factor, L(x) is the Langevin function, $x = \mu H_{FC}/k_B T_f$, $\mu_B = 9.274 \cdot 10^{-24}$ J/10 kOe is the Bohr magneton, $k_B = 1.381 \cdot 10^{-23}$ J/K is the Boltzmann constant and $\mu$ is the magnetic moment of the clusters, $H_{FC}$ is cooling field strength and the measured temperature, $T_f = 5$ K. From the Fig. 2d, it is evident that the experimental data could be well described by the above relation as shown by solid red curve. The obtained value for the $J_i$ for PCMO10 ~ - 10 meV. Negative interface exchange constant represents the AFM coupling between the FM domain and AFM phase. The $J_i$ value for PCMO10 is within the range that has been reported for other manganites $Pr_{1/3}Ca_{2/3}MnO_3$ [29] and cobaltite $La_{0.82}Sr_{0.18}CoO_3$ [30]. In contrast, the $H_C$ is found to increase and saturate with increase in $H_{FC}$, as shown in Fig. 2d, typical for EB systems[14].

To investigate the training effect[31] of the EB, a series of MH loops were consecutively measured at 5 K after FC. The ZFC and FC loops are shown in the main panel of Fig. 3. FC essentially establishes a shift in the first (n = 1) MH loop. The $H_{EB}$ values for the subsequent MH loops are reduced significantly due to the collinear arrangement after the first magnetization reversal of the ferromagnet.[5] Empirically, the training effect can be quantified by a power law function[32] $H_{EB}(n) = H_{EB}^\infty + Dn^{-\alpha}$, where $H_{EB}^\infty$ is the limiting value of $H_{EB}$, when the number of cycles n approaches infinity, D is a constant and $\alpha$ is a positive exponent whose best fitting value is about 0.35. The bottom right inset of Fig. 3 shows the variation of $H_{EB}$ with n after FC and the solid red curve shows the best fitting result for n > 1. The obtained fitting parameters, $H_{EB}^\infty$ and $\alpha$ are 513 Oe and 0.35 respectively. Apparently, power - law cannot explain the steep training effect between the first and the second MH loops. However, we find that the training effect could not be described well (fitting is not shown) with the well-known spin configuration relaxation model as well for n ≥ 1.

In summary we have investigated the exchange bias and training effects in PCMO10. It has been demonstrated that the exchange bias effects are due to competing interactions between FM/AFM phases. Temperature dependent oscillatory exchange bias is observed below SG freezing temperature and this phenomenon may be attributed to ICDW present in PCMO10. The reduction in the $H_{EB}$ is observed after cycling the system through several consecutive hysteresis loops and the dependence of $H_{EB}$ on 'n' has well been described by power-law behavior for $n > 1$.

SNJ would like to thank KU Leuven for research fellowship. This work is supported by the K.U. Leuven Excellence financing (INPAC), by the Flemish Methusalem financing and by the IAP network of the Belgian Government. The authors thank the reviewer for providing valuable suggestions.

Corresponding author:

Surya.Jammalamadaka@fys.kuleuven.be, srinivasasingamaneni@boisestate.edu

**Figures**

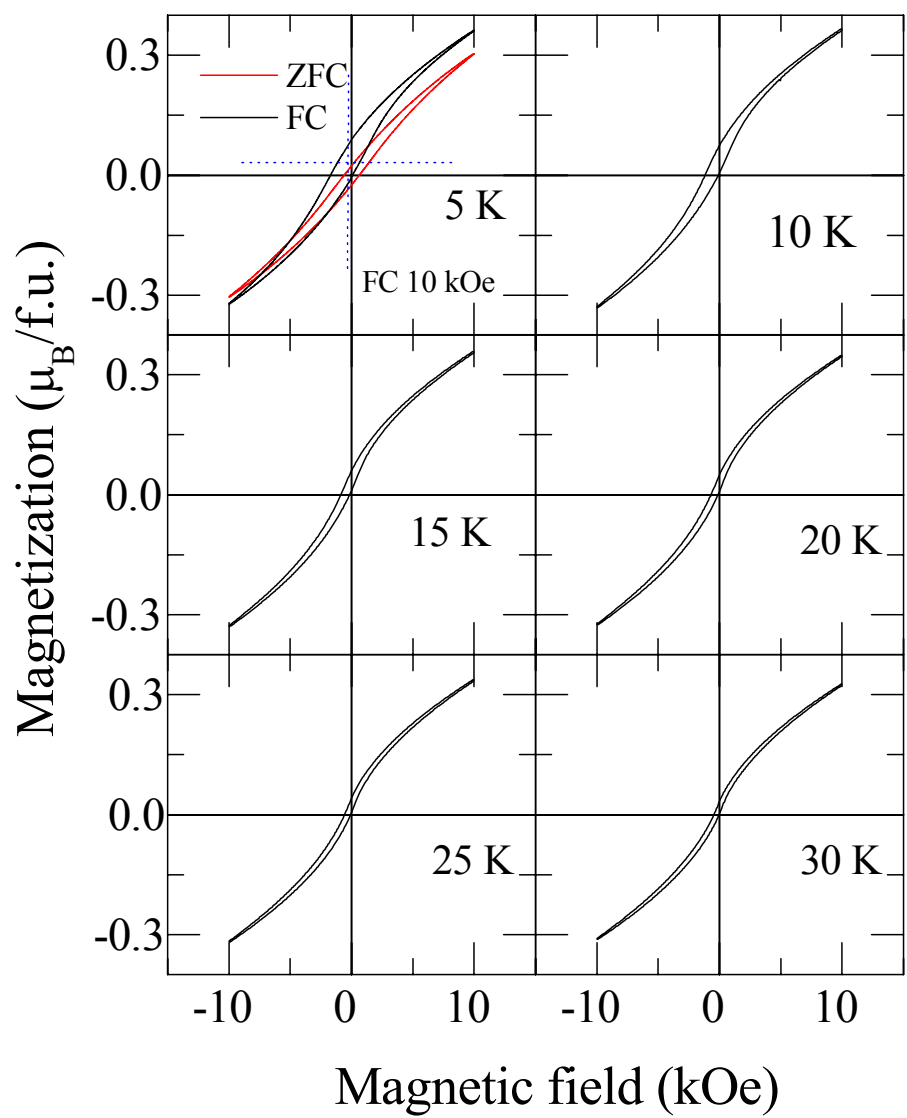

**Fig. 1:** Isothermal M-H curves measured under ZFC conditions at different T's for PCMO10. The horizontal and vertical loop shifts can be clearly noticed from these curves. To show that the FC causes the vertical as well as horizontal loop shift, a FC MH loop has also been shown at 5 K.

As a representative figure, we also displayed field cooled (FC) MH loop at 5 K measured under field-cooling strength of 10 kOe.

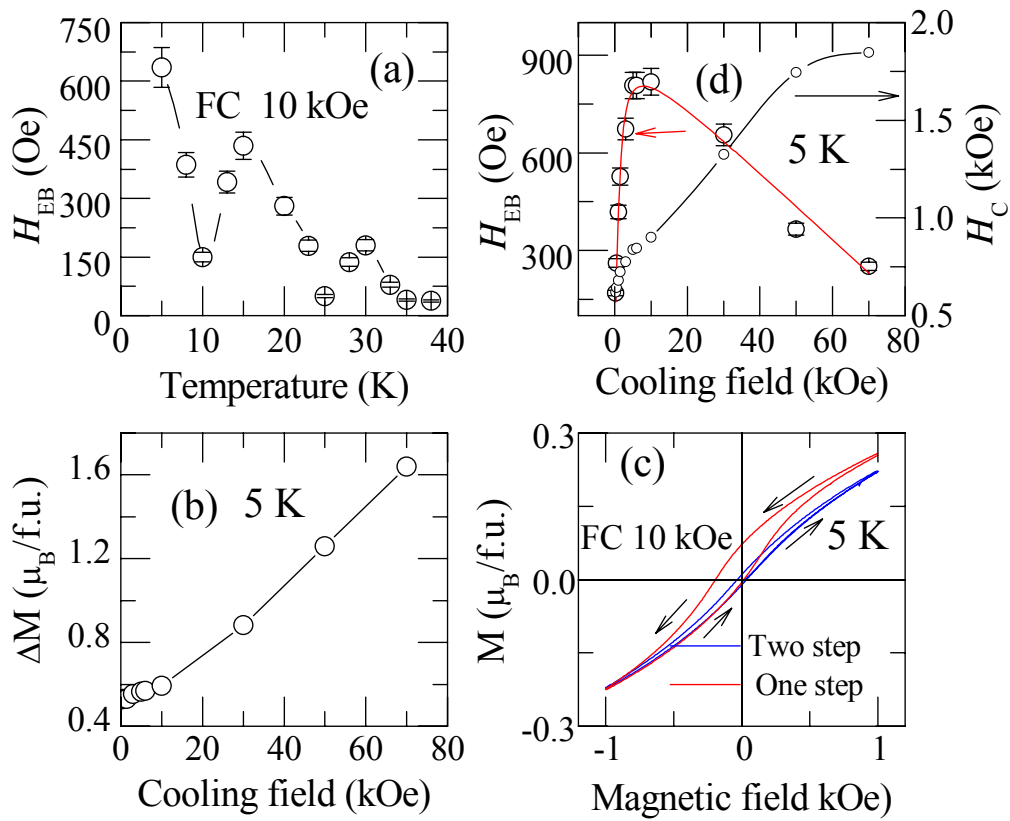

**Fig. 2:** (a) T variation of oscillatory exchange bias field (OEB) (b) Variation of ΔM with FC (c) M-H loops resulted from one step and two step field cooling process. Symmetric MH loops are evident after two step process (d) $H_{EB}$ and $H_C$ variation with FC. The red curve is the fit obtained from equation (1), scattered points are the experimental data.

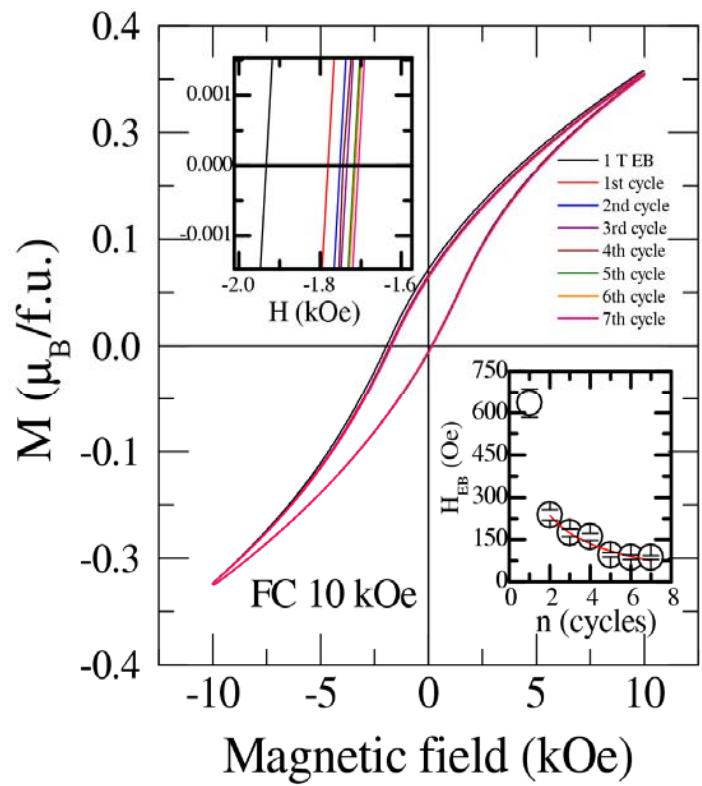

**Fig. 3:** (Color online) Consecutively measured seven MH loops after field cooling PCMO10 in the presence of 10 kOe. Top left inset shows the enlarged view to reveal the shift in the MH loop to the lower fields as the loop number (*n*) increases. Bottom right inset shows the variation of $H_{EB}$ on 'n' at 5 K after FC in 10 kOe. The solid line illustrates the best fitting with the power law for $n \geq 2$. The obtained fitting parameters are $H_{EB}^{\infty}$ = 513 Oe and α = 0.35.